\documentclass{article}
\usepackage{psfig}
\usepackage{sao2}
\issue{2005, 58, 5-15}
\begin{document}
\markboth{Verkhodanov, Parijskij, Starobinsky}
{Determination of $\Omega_\Lambda$   and $H_0$...}
\title{
Determination of $\Omega_\Lambda$   and $H_0$ from photometric data of
radiogalaxies
}

\author{
O.V. Verkhodanov\inst{1}
\and Yu.N. Parijskij\inst{1}
\and A.A. Starobinsky\inst{2}
}

\institute{
\saoname
\and Landau Institute of Theoretical Physics, Moscow
}
\date{July 23, 2004}{September 15, 2004}
\maketitle

\begin{abstract}
    From photometric observations of elliptical galaxies, among which
are both
radio galaxies and radio-quiet objects, an investigation was carried out
of the relationship `redshift -- age of the stellar system'
$(\Delta z/\Delta t)$.
By means of  this relationship cosmological parameters $H(z)$ and
$\Omega_\Lambda$ are estimated. Ages of stellar systems are determined
within the framework of evolution models of synthetic spectra PEGASE and
GISSEL. This approach can be considered as time study of objects of the
early Universe independent of other cosmological models. Construction of a
pooled sample is described, containing 220 objects from different populations of elliptical
galaxies, for which an analysis of the upper limit of the age of formation of
a stellar system was performed. These data were used to estimate the
boundaries of determination of the cosmological parameters $H_0$ and
$\Lambda$--term: $H_0=72\pm10$ and $\Omega_\Lambda=0.8\pm0.1$ in the model
GISSEL and $H_0=53\pm10$, and $\Omega_\Lambda=0.8\pm0.1$ in the model PEGASE.

\end{abstract}

\section{Introduction}

An essential point in determining cosmological parameters is the independence
of the procedure of the classical methods, which have come to be classical,
such as deep three-dimensional surveys of galaxies and clusters,
1a--type supernovae (see, for instance, Leibundgut 2001) or relic radiation
(for instance Efstathiou et al., 2002).

One of the independent techniques is based on the datings associated with the age
of  galaxies (for instance, Saini et al., 2000).
The first attempts to estimate
$\Omega_\Lambda$ with the use of ages stellar systems were made a few years
ago (see, for instance, Parijskij, 2001) proposed by Jimenez and Loeb (2002).
It is based on the datings connected with variations of ages of galaxies
determined by the spectroscopic technique.

Such an approach makes it possible to construct an independent chronological
scale applicable to the early stages of evolution of the Universe. This
approach is based on measurements of differences of the ages $Delta t$
between two passively evolving galaxies which form at the same time but
separated by a small interval $Delta z$. Then one can determine the
finite difference $(\Delta z/\Delta t)\approx dz/dt$. All the galaxies
in the procedure proposed by Jimenez and Loeb must have alike metallicities
and low rates of star formation (i.e. red colors), while the mean age of
the system must be considerably larger than the difference of ages of
galaxies $\Delta t$. By applying this differential method, Jimenez and Loeb
(2002) suggest H(z) and $\omega_Q(z)$ to be measured directly from the first
and second derivatives $(\Delta z/\Delta t)$ $(\Delta^2 z/\Delta t^2)$:

\begin{equation}
H(z)= -{1\over (1+z)}{dz\over dt}\,,
\label{eq:H_z}
\end{equation}

\begin{eqnarray}
H_0^{-2}\frac{d^2z}{dt^2}=\frac{[H_0^{-1}({dz}/{dt})]^2}{(1+z)}\left
  [\frac{5}{2}+\frac{3}{2}\omega_Q(z)
  \right]-\nonumber \\
\frac{3}{2}\Omega_m(0)(1+z)^4\omega_Q(z)\,.
\label{eq2}
\end{eqnarray}

By proposing this differential method, the authors show that it is necessary
both to increase the sample of galaxies and to improve the signal/noise
ratio.

We have used a similar method but with another  type of dating of
the age, namely, from photometric data and choosing optimum consistency
of the distribution of energy in the spectrum (SED), depending on
age, by the observed fluxes. This procedure, which has already
become standard, operates with sufficient stability for a pure sample
of elliptical galaxies (see, for instance, Verkhodanov et al., 1999),
though it may give an error in the age up to 2 Gyr.

The methods using color and spectral ages of galaxies are based on
chronometrying the rates of expansion of the Universe from physical
processes not associated with  cosmology: from the rates of nuclear
reactions in stars, the knowledge of which for standard stars of
the solar type
are accurate enough and in the last decade have obtained
numerous direct and indirect confirmations, including the latest achievements
of acoustic tomography of the entrails of the Sun. For this reason, the
proposed method of chronometrying the evolution of the Universe resembles
those of chronometrying on Earth
from the data of radioactive decay and in any case is independent of others.

Note that giant elliptical galaxies with high radio luminosity and with the
old stellar population are the most suitable objects
for estimating the age of stellar
systems. The present-day models predict fast enough (during 1 billion years)
formation of such systems at $z\sim4$ (Pipino \& Mantencci 2004), which
enables application of photometric methods to their investigation. The
efficiency of selecting such galaxies with the aid of radio astronomy
methods, beginning from moderate redshifts ($z>0.5$) is confirmed by several
groups (Pedani 2003). A combined diagram of Hubble ``K--z''
for radio galaxies
and field galaxies (Jarvis et al. 2001; De Breauk et al. 2002) show that
radio galaxies have the highest luminosity at any redshift $0<z<5.2$
(Reuland et al., 2003). Besides, radio galaxies have supermassive black
holes whose mass is generally  proportional to a stellar bulge one
($M_{BH}\sim0.006M_{buldge}$,
Maggorrian et al. 1998), and this fact is additional evidence of the
presence of an already formed stellar population. Note that the estimate of
the age of distant galaxies is also of interest in connection with searching
for primeval black holes with masses
$10^3-10^6 M_{\sun}$.

Formation of radio galaxies at redshifts $z\sim3-5$
provide the already formed
stellar populations at $z\sim2-4$ in the $\Lambda$CDM models.
Thus, when selecting distant
radio galaxies, we isolate with sufficient efficiency giant elliptical
galaxies which can be used to estimate the age of a stellar
population and to investigate the ratio $t(z)$ (Parijskij 2001;
Verkhodanov \& Parijskij 2003, Starobinsky et al. 2004).

The present paper describes approaches, methods and results concerning
cosmological parameters estimates using samples of elliptical galaxies.
We will discuss, first, the problems connected with the use of evolutionary
synthetic models of spectra of galaxies. Then, using the data on the
``Big Trio'' project
(Parijskij et al. 1996, 2000a, 2000b; Parijskij 2001; Soboleva
et al. 2000; Kopylov et al. 1995; Verkhodanov et al. 2002) and other
authors (Verkhodanov et al. 1999) we will present results of the first
attempts of correction of the standard cosmological model with cold dark
matter (SCDM) by means of age estimates of parent galaxies responsible
for the origin of powerful radio galaxies at large redshifts. Further we
will give a summary of attempts to estimate the relationship between
the age of
galaxies and their redshifts from the current evolutionary  models of the
stellar population of elliptical galaxies for a wider interval of redshifts,
including close, $z<1$, objects.

\section{Photometric dating}

\subsection{Evolutionary models of spectra}

At the end of the 1980s and in the early 1990s attempts were made to employ
color characteristics of radio galaxies to estimate redshifts and ages of
stellar systems of parent galaxies. Numerous evolutionary models were
proposed, which led to result strongly different
from one another in comparing
with observational data (Arimoto \& Yoshii 1987; Chambers \& Charlot 1990;
Lilly 1987,1990; Parijskij et al. 1996).
Over the past few years the models PEGASE: Project de'Etude des Galaxies par
Synthese Evolutive (Fioc \& Rocca-Volmerange, 1997) and GISSEL'98: Galaxy
Isochrone Syntheses Spectral Evolution Library (Bruzual \& Charlot 1993;
Bolzonella et al., 2000), in which the defects of previous models were
eliminated, have been widely used.

In the ``Big Trio'' experiment (Parijskij et al. 1996) we also attempted
to apply these methods to distant objects of the RC catalog with steep
spectra, for which
we measured the values in the four filters (BVRI). The procedure of smoothing
was used, which
made it possible to simulate and predict the flux in
the given filter with the given SED with allowance made for
the filter response function
of this band
and also with the effects of the
redshift allowed for. These changes in the procedure permitted the results
to be more reliable in comparison with the previous paper work.

Preliminarily (Verkhodanov et al. 1999) we investigated applicability of new
models to populations of distant ($z>1$) radio galaxies with
the known redshifts,
for which we have managed to find in the literature more or less reliable
data of multicolor photometry in the optical and near-infrared
not less than in three filters. In particular, it is shown that redshifts
can be estimated to an accuracy $25-30\%$ at $1<z<4$, given the measured
stellar magnitudes in more than three filters. But if, at least, one
brightness estimate in the infrared range is available, than it suffices
to use measurements in the tree filters. Estimations were made for two
evolutionary models PEGASE (Fioc \& Rocca-Volmerange 1997), which
was constructed for the galaxies of the Hubble sequence both with
star formation and passively evolving.
One of the advantages of this model consists in the extension to the
near-IR (NIR) range of the atlas of the synthetic spectra of
Rocca-Volmerange and Guiderdoni 1988).
This model reconsiders a library of stellar spectra
which is computed with allowance
made for parameters of cold
stars. The model covers a range from 220\AA~ to 5 microns. According to the
authors, the algorithm of the model traces rapid evolutionary phases,
such as those of the  red supergiant or AGB in the near-IR range.
For the computation a wide set of SED curves was used for massive
elliptical galaxies in a range of ages $7\times10^6$  years to
$19\times 10^9$ years.  We have also used the computations for the
elliptical galaxies of the library of synthetic spectra of the model
GISSEL'98 (Bolzonella et al. 2000). The spectra are constructed with the
aid of the evolutionary models of Bruzual and Charlot (1993,1996). For the
calculation of the synthetic spectra of the elliptical galaxies of this
library, the following parameters of star formation were assigned:
simple star formation (SSP -- simple stellar population), the duration  of
the process of star formation is 1 billion years,
while decaying of the burst activity of star formation proceeds by an
exponential law. The model used solar metallicity. The initial mass function
(IMF) with an upper limit of 125 solar masses has been taken from the paper by
Miller and Scalo (1979).
As is shown in the paper by Bolzonella et al. (2000), the choice of the IMF
does not effect the accuracy of determination of redshifts. The model
tracks are calculated in a wavelength range from 200 to 95800\AA.
For our computations we have used the range assigned by a redshift limit
from 0 to 6.
The sets of evolutionary models are accessible at the server
{\tt http://sed.sao.ru} (Verkhodanov et al., 2000).

\subsection{ Procedure of estimating the age and redshift}

Prior to the application of model curves we carried out their smoothing by
the filters with the application of the following algorithm (Verkhodanov et
al., 2002):
%
%
\begin{equation}
  S_{ik} = \frac {\sum\limits_{j=1}^n s_{i-n/2+j} f_{jk}(z)}
		 {\sum\limits_{j=1}^n f_{jk}(z)},
\end{equation}
where $S_i$ is the initial model SED curve,
$S_{ik}$ is the one smoothed by the $k$-th, $f_k$(z) is the curve of
transmission
of the $k$-th filter ``compressed'' $(1+z)$ times
when moving along the axies of the
point in the filter response function.
From the $k$ curves of SED
thus formed, we have constructed a two-dimension array ($\lambda$--filter)
of smoothed synthetic stellar spectra for further computations.

The estimation of ages and redshifts of radio galaxies was made by the method
choosing on the smoothing curves of SED of optimum positions of photometric
values obtained in different bands in the observations of galaxies. We
have used SED curves already computed and stored in tables
for different ages. The algorithm of the choice of optimum position of
points on the curve
consisted
(Verkhodanov 1996) in shifting the observational points along the SED curves.
In so doing, such a position was found at which the sum of the squares of
the departures of the points from the corresponding smoothed curves is
minimum, i.e. the minimum of $\chi^2$ was actually computed
\begin{equation}
\chi^2 = \sum\limits_{k=1}^{N filters}
      \left( \frac{F_{obs,k} - p\cdot {\tt SED}_{k}(z)}{\sigma_k}\right),
\end{equation}
where $F_{obs}$,
$k$ is the observed stellar magnitude in the $k$-th filter, ${\tt SED}_k(z)$
is the model stellar magnitude for the given spectral distribution in the
$k$-th filter at the given $z$, $p$ is the free coefficient, $\sigma_k$
is the measurement error. The redshift was determined from the shift of
the position of the observed magnitudes at their best position on the SED
curves from the position ``rest frame''. From the general
set of curves we chose such ones on which the sum of the squares of
discrepancies for the given observations of radio galaxies prove to be
minimum.

We checked the correctness of estimates of ages (and redshifts) by
2 methods. In the first one we took synthetic spectra obtained by means of
smoothing by filters the SED curves for different age. This procedure
made it possible to simulate CCD observations
for 5 filters. Further the points were chosen corresponding to
the filters VIJHR, for instance, at the redshift $z=0.54$ and also the model
GISSEL with SEDs for 1015.1 and 5000~Myr. Two tests were applied for
each age to estimate the value: with fixed $z=0.54$ and unfixed redshift.
From the results of testing a conclusion can be drawn that the age and
redshift are defined reliably, however, a falling on neighboring curve of ages
is possible, which gives an error of 200~Myr, while at unfixed $z$ the
result is also affected the discretization in wavelength $\lambda$ in
the SED curves (the error in $z$ is up to 60\%).

In the second case capabilities were studied of the method of determination
of the redshifts and ages of the stellar population of parent galaxies from
the data of multicolor photometry.
For this purpose, we have selected about 40
distant galaxies with known redshifts, for which stellar magnitudes in not
fewer than four filters (Verkhodanov et al., 1998b, 1999)
are presented in the literature.
At first, using the selected
photometric data with the use of the models PEGASE and GISSEL'98 only ages
of the stellar population of parent galaxies at a fixed known redshift were
determined. Then a search was made for an optimum model of the SED curve with
a simultaneous determination of the redshift and age of the stellar
population. After that a comparison of the obtained values was made.
By this method we estimated both the age of the galaxy and the redshift within
the framework of the given models (see also Verkhodanov et al. 1998a,
1999). It is clear from general considerations that the reliability of the
result at large redshifts is strongly affected by the presence of infrared
data (up to the K range) because when fitting we overlap the region of rapid
jump of the spectrum before the region of SED, and thereby we
can reliably, with a well defined maximum on the plausibility curve,
determine the position of our data. Indeed, when checking the reliability
of the procedure with the aid of the measurements available with keeping
of only 3 points, one of which is in the K range, we obtain the same result
on the plausibility function as from 4 or 5 points. If the infrared range is
not used, then the result turns out to be more uncertain. However, as it is
shown in the paper by Verkhodanov et al. (1999), the variant of  4
filters close
disposition  as in our case of BVRI photometry yielded a good result in
the sample of 6
objects. This result  coincides with the one obtained when all the filters
were used, including the infrared range.

\section{Sample of objects}

It should be noted that the choice of elliptical galaxies as objects for our
investigation is not accidental. They can be considered as the most optimum
objects among stellar systems having a homogeneous enough stellar population.
Although such objects have (moderate) metallicity gradients
(Friaca \& Terlevich 1998),
the modeling showed (Jimenez,
Loeb, 2002) that the variation of metallicity leads to an
uncertainty of estimates of the age within 0.1 Gyr, which lies inside
uncertainties of estimates.

In the given investigation we use radio galaxies which, as a rule, are
identified with giant elliptical galaxies (gE) and are good ``lanterns'' and
representatives of distant stellar systems. The standard point of view of the
last decades has been that powerful radio galaxies are associated with old
huge stellar system of the gE-type having the red color. The experience of
using globular clusters in our Galaxy to estimate the age of the Universe
shows that the search for the oldest stellar systems at large redshifts
may be useful for chronometrying of the rates of expansion of the Universe
at any distances  at which powerful radio galaxies still existed. As many
groups (Rawlings et al. 1996; van Beugel et al 1999), including the
project ``Big Trio'' (Soboleva et al. 2000) have shown that powerful
galaxies appeared at redshifts of about 5. The whole interval $0<z<5$
can be potentially investigated even today since the sensitivity of radio
and optical telescopes is sufficient for investigation of such powerful
radio and optical
objects. In contrast to quasars, the radiation of the stellar
population can be readily isolated from that of the gaseous component.
Note, however, that in radio galaxies uncertainties arise in photometric
determination of the age because of different factors (see, for example,
Moy \& Rocca-Volmerange, 2002), such as ionization and transillumination
of the radiation from the nucleus, interaction of clouds and jets etc.
Besides, the galaxies at early stages could be interacting, which
changes the stellar
population. Nevertheless, the radio galaxies remain so far the only simple
means of studying elliptical galaxies at large redshifts.

\subsection {Data on radio galaxies from the catalog ``Cold''}

The given sample is composed from FRII-type galaxies found in the RATAN-600
survey ``Cold'' (Parijskij et al., 1991, 1992) with involvement of data of
multicolor photometry for estimating color redshifts and ages of stellar
systems of parent galaxies (Parijskij et al. 1996; Verkhodanov et al. 2002).
Later, spectral observations at BTA with the device SCORPIO (Afanasiev et al.
2002) were carried out, which confirmed with high accuracy (correlation
coefficient 0.92) the photometric estimates.

In the program ``Big Trio'' BVRI values of about 60 radio galaxies were
estimated,
and it was
discovered that although the color age does not have a large dispersion, the
upper limit of the age is a sufficiently reliable function of
redshift (the larger z, the less the maximum age). A comparison of this
upper limit with the SCDM model showed that age is not at variance with the
SCDM model without the cosmological constant $\Lambda$, but in the interval
$0.7<z<2$ there are objects with the color age greater than the age of the
Universe at the corresponding redshift. Such a situation, as it is known
(see, for instance, Sahni and Starobinsky 2000) is eliminated in the
$\Lambda$CDM models.
Indeed, the age of such a Universe does not differ from the SCDM
model either at very small or at very large redshifts, which is seen from
the formulae presented in the paper mentioned.

However, in the interval of redshifts $z=1-2$ the difference may reach 1--2
billion years, which is close to the possibilities of the experiment. The
simple theory $\Lambda\not=0$ for a spatially flat isotopic cosmological
model yields a
relationship between the position of a maximum of departures from the SCDM
model on the axis of redshifts and the value of cosmological constant.
The first attempts to estimate the value of the cosmological constant from
the ``Big Trio'' data were made in 1999 (Parijskij 2001). In connection with
the uncertainty in quantitative estimates of the age of galaxies
from the measured redshifts, percentage
of galaxies whose age formally exceeds that of the Universe in the SCDM
 model with $\Lambda=0$
 was estimated. Then a histogram of distribution of number of these galaxies
as a function of $z$
 was constructed (Fig.\,1).
\begin{figure*}[!th]
\centerline{
\psfig{figure=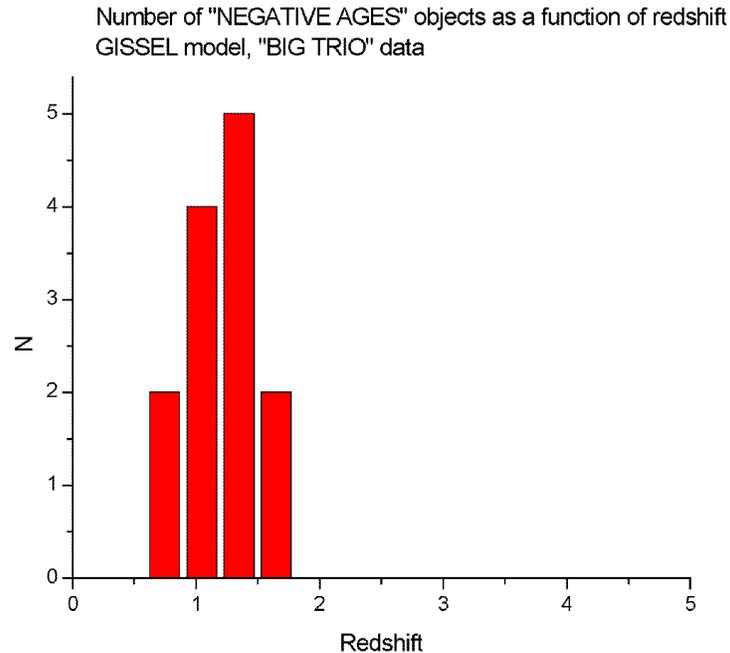,width=12cm,angle=-90,bbllx=-196pt,bblly=-218pt,bburx=793pt,bbury=1062pt,clip=}
}
\caption{Histogram of the distribution of galaxies in $z$
with a formal age above that of the Universe.}
\end{figure*}

\begin{figure*}[!th]
\centerline{
\psfig{figure=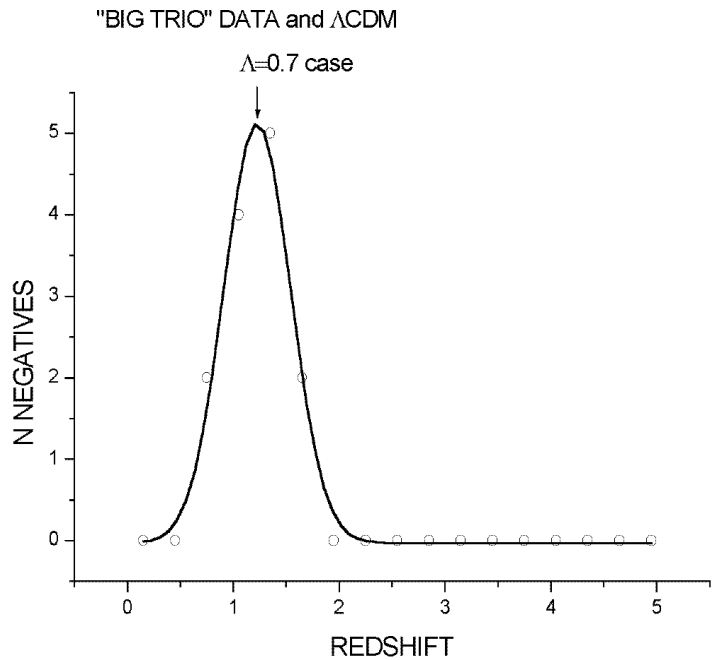,width=12cm,angle=-90,bbllx=-196pt,bblly=-218pt,bburx=793pt,bbury=1062pt,clip=}
}
\caption{Estimates of $\Omega_{\Lambda}$ from the objects of the project
 ``Big Trio''.}
\end{figure*}
\begin{figure*}[!th]
\centerline{
\psfig{figure=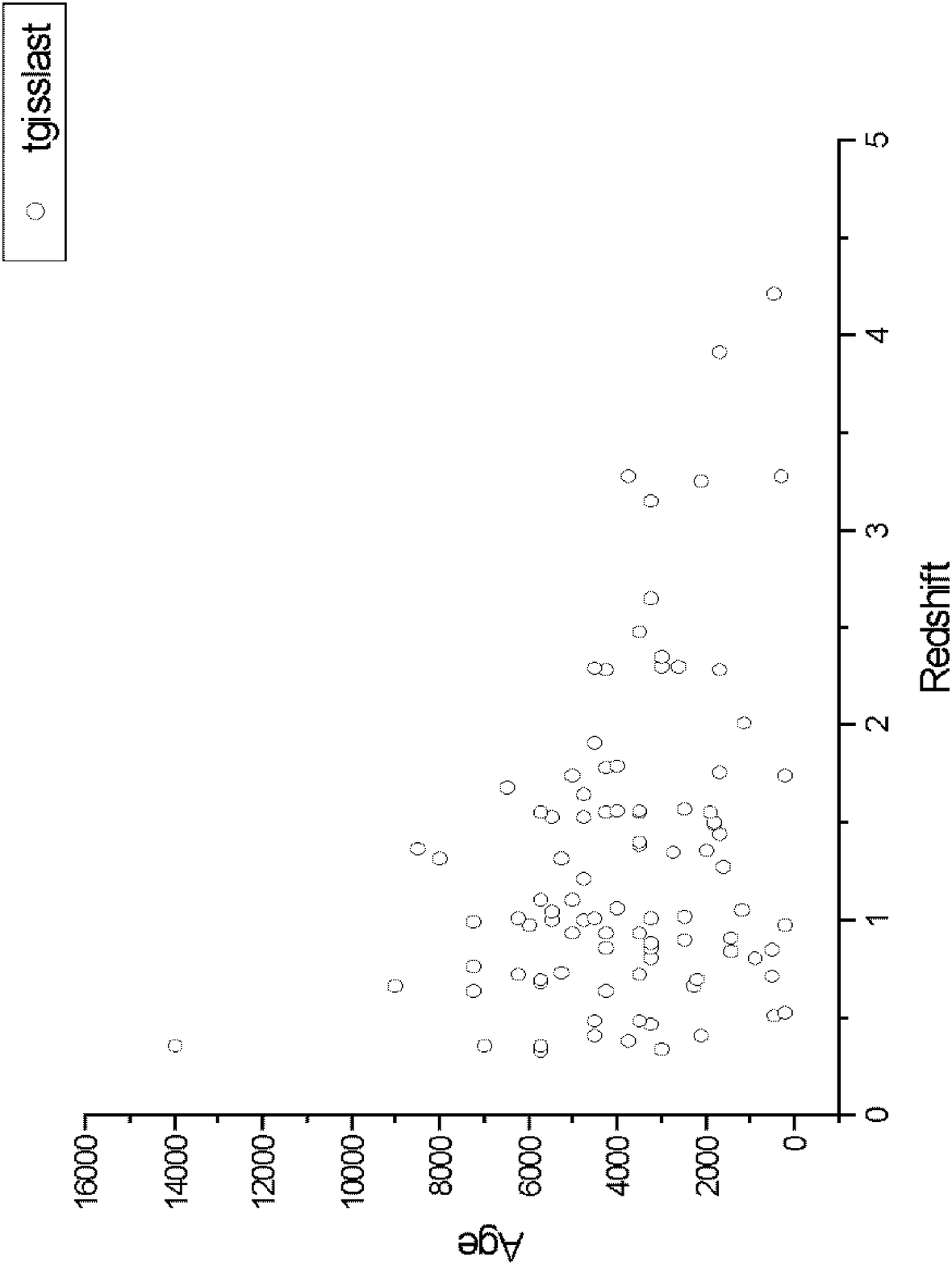,width=12cm,angle=-90,bbllx=-196pt,bblly=-218pt,bburx=793pt,bbury=1062pt,clip=}
}
\caption{
Relationship $t(z)$ for radio galaxies having steep radio spectra
with large $z$ taken from different published papers.
It is seen that at $z<2$  the dispersion of ages is large.
}
\end{figure*}
From the position of the maximum the proportion of
``dark matter'' ($\Omega_\Lambda$)
which turned out to be close to the value derived
from 1a--type supenovae was estimated: $\Omega_\Lambda=0.8-0.6$ (Fig.\,2).
This result stimulated our further steps in the usage of age characteristics
of the stellar population. Fig.\,3 shows all the data collected by the group
``Big Trio'' in 2001 from steep spectrum radio galaxies.
It is seen that there
present objects with large $z$, but at $z<2$ dispersion of ages is great.
But the
larger the redshift, the less the age of the oldest object, as it was to be
expected in all evolutionary models of the Universe. Having chosen a
population of objects the age of which is younger than that of the Universe in
the $\Lambda$CDM model by more than 2 billion years,
obtain a relationship $z(t)$
similar to the one displayed in Fig.\,4
\begin{figure*}[!th]
\centerline{
\psfig{figure=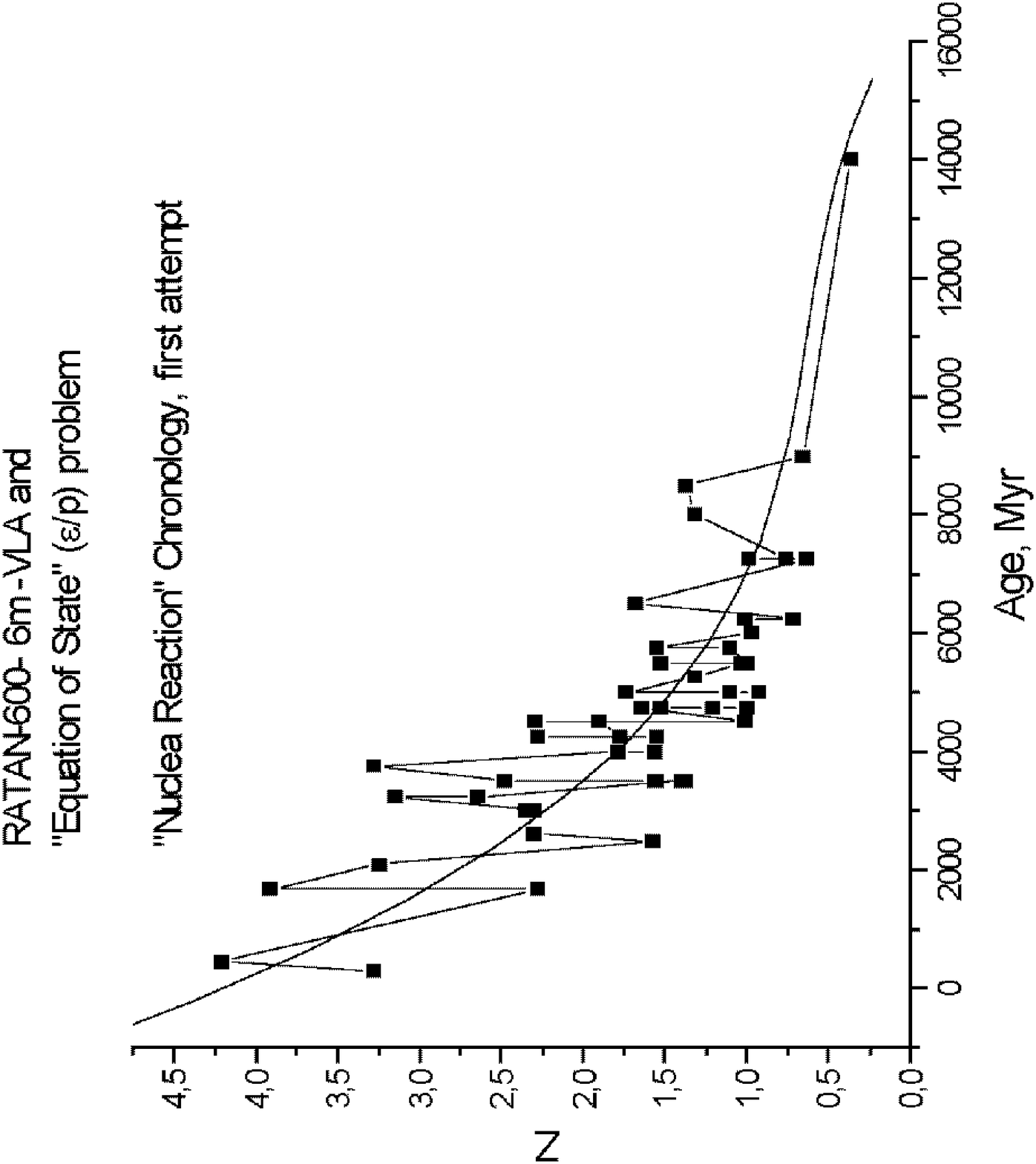,width=12cm,angle=-90,bbllx=-196pt,bblly=-218pt,bburx=793pt,bbury=1062pt,clip=}
}
\caption{Relationship $z(t)$ of objects with
$t_{star\,form}-t_{Univer}\le$ 2\,billion years}
\end{figure*}
The relationship of such a type can already serve as a basis for estimating
$R(t)$.

\subsection{Data on investigated radio galaxies with $z>1$}

As it was stated above, to check the procedure and estimate  redshifts
and age of stellar systems, we made a sample of radio galaxies FRII with
redshifts up to $z=3.80$ (Verkhodanov et al. 1998b, 1999) using data obtained
by other authors.

It should be noted that literature photometric data are very inhomogeneous.
They were obtained not only by different authors but also with the help of
 different
instruments with different filters. It was not always that measurements for
one  and the same object were made with equal apertures etc. For this
reason, after the final selection from 300 radio galaxies of the initial
sample only 42 remained. The greater part of the sample turned out to be
beyond the limits of the sample because they have the properties of quasars,
which impedes strongly the use of the procedure SED for standard elliptical
galaxies
{\small
\begin{table*}[!th]
\caption{Selected elliptical galaxies --- members of clusters.
The table contains the following
fields: equatorial coordinates and name of the cluster, numbers of selected
galaxies from the cluster according to the published order (Stanford
et al. 2002), redshift, used filters, K--limit}
\begin{center}
\begin{tabular}{|cllclc|}
\hline
 RA+Dec(2000)     &   Name          & ID numbers of galaxies    &  $z$  & filters & K$_{lim}$  \\
\hline
001631.2$+$791649 & 3C 6.1          & 14,18,33,34,43                 & 0.840  & KJIR   & 19.8  \\
001833.5$+$162515 & Cl 0016+16      & 8,9,15,17,22,27,35,40,42       & 0.545  & KHJIV  & 19.1  \\
002354.5$+$042313 & GHO 0021+0406   & 9,12,15,24,29                  & 0.832  & HJIR   & 20.0  \\
002635.7$+$170945 & Cl 0024+16      & 3,4,5,8,9,10,16,26,33          & 0.391  & KHJRg  & 18.8  \\
004910.9$-$244043 & Vidal 14        & 6,11,14,17,23,31,40,43         & 0.520  & KJIV   & 18.0  \\
005657.1$-$274030 & J1888.16CL      & 7,17,23,45                     & 0.560  & KHJIV  & 19.2  \\
011018.5$+$314719 & 3C 34           & 7,8,13,21,23,34,40             & 0.689  & KHJiV  & 19.1  \\
030619.1$+$171849 & GHO 0303+1706   & 6,7,13,18,22,24,34,40          & 0.418  & KHJRg  & 18.8  \\
032001.4$+$153200 & GHO 0317+1521   & 3,8,9,13,14,23,24              & 0.583  & KJIV   & 19.2  \\
041246.6$-$655055 & F1557.19TC      & 20,25,30,37,39                 & 0.510  & KHJIV  & 19.1  \\
045410.9$-$030057 & MS 0451.6-0306  & 3,11,12,19,25,31,33,40         & 0.539  & KHJiV  & 19.2  \\
073924.3$+$702315 & 3C 184          & 3,4,8,12,15                    & 0.996  & KJI    & 20.3  \\
084835.9$+$445337 & RDCS 0848+4453  & 4,6,9,11,13,15                 & 1.273  & KHJIR  & 20.5  \\
085809.9$+$275052 & 3C 210          & 3,6,13,15,16,17                & 1.169  & KJI    & 20.5  \\
093239.6$+$790632 & 3C 220.1        & 5,8,12,16,17,19,24,25          & 0.620  & KHJIV  & 19.5  \\
105659.5$-$033736 & MS 1054.5-032   & 4,7,9,13,14,22,23,25,26,30     & 0.828  & KHJiR  & 20.3  \\
114022.2$+$660813 & MS 1137.5+6625  & 3,8,10,11,12,15,16,17,20,21,24 & 0.782  & KHJiR  & 20.0  \\
132448.9$+$301138 & GHO 1322+3027   & 2,5,8,10,17,18,20              & 0.751  & KHJiR  & 20.3  \\
141120.5$+$521210 & 3C 295          & 9,18,24,27,28,31,32,34,35,37   & 0.461  & KHJiV  & 18.8  \\
151100.0$+$075150 & 3C 313          & 6,12,13,21,33                  & 0.461  & KJiV   & 18.5  \\
160312.2$+$424525 & GHO 1601+4253   & 4,6,17,18,19,22,43             & 0.539  & KHJiV  & 19.2  \\
160424.5$+$430440 & GHO 1603+4313   & 5,15,17,36,40                  & 0.895  & KHJiR  & 20.3  \\
160436.0$+$432106 & GHO 1604+4329   & 7,10,20,21,28,30,32            & 0.920  & KHJiR  & 20.1  \\
205621.2$-$043753 & MS 2053.7-0449  & 35,39,51,89                    & 0.582  & KJIV   & 19.2  \\
220403.9$+$031248 & GHO 2201+0258   & 10,11,13,14,17,25,29,35,38     & 0.640  & KJIV   & 19.3  \\
\hline
\end{tabular}
\end{center}
\end{table*}
}
 The derived
median value of the age for the given sample is 5\,Gyr for the model GISSEL
and 9\,Gyr for the model PEGASE.

\subsection{Clusters of galaxies}

The subsample of elliptical galaxies from clusters, which was proposed by
A.\,Kopylov (2001) is the most representative from the investigated group of
objects (Table 1). We have used for its compilation the data from the paper by
Stanford et al. (2002) containing a sample of 45 clusters of galaxies at
 redshifts $0.1<z<1.3$.
For all the objects photometric data from the optical and near-infrared
region are available. On the average, stellar magnitudes in the bands
BIJHK are presented for each galaxy.
For our sample we have selected by the color index 5--9 objects, typical
elliptical galaxies, from 25 clusters, a total of 175 elliptical galaxies.
Table\,1 presents the selected clusters with the numbers of galaxies which
photometric data are used to estimate
the age of stellar systems. The bands in which observations of clusters were
carried out, their redshifts and K-values are also presented.

\section{Procedure of estimating parameters}

Our approach is based on the analysis of the function $t(z)$:
%
\begin{equation}
t(z) = \int\limits^\infty_z \frac{d\mbox{\~z}}{(1+\mbox{\~z})H(\mbox{\~z})},
\end{equation}
constructed from ages of radio galaxies depending on the redshift. As the
function $H(z)$ we used the expression
\begin{equation}
H^2 = H^2_0 [ \Omega_m (1+z)^3 + A + B (1+z) + C (1+z)^2 ],
\end{equation}
where $A+B+C=1-\Omega_m$.

The fitting of function $t(z)$ to the data analyzed was performed with
the aid of variation of four parameters ($H_0$, $\Omega_m$, $A$, $B$).
We divided
the whole set of redshifts into equal intervals  $\Delta z$  and used
the maximum age value in each of the intervals. From the sum of the squares
of discrepancies a four-parameter plausibility was constructed. With the
values of the parameters $B=C=0$,
i.e. when the simplified model of the function
$H(z)$ defined only by two parameters ($H_0$, $\Omega_m$) and
$A= 1-\Omega_m=\Omega_\Lambda$ was used, there are stable solution
of both models of evolution of the stellar population. The results of
determination of the parameters are listed in Table\,2, in which the
parameters of approximation of the curve for the intervals $\Delta z=0.2$
and 0.3 are given for also in Fig\,.5.
\begin{table*}[!th]
\caption{
A two-parameter fitting of cosmological parameters by formulae (5) and (6)
(at $B=C=0.0$) for the approximation curves in the interval
$\Delta z=0.2$ and
0.3 for both models of stellar population.
In the columns are presented: the used model of the stellar population, the
interval $\Omega_m$, $\Omega_\Lambda$, $H_0$, the discrepancy $\epsilon$
of the relationship $t(z)$, the relative discrepancy $\epsilon/T_0$,
where $T_0$=13.7\,Gyr is the age of the Universe.
}
\begin{center}
\begin{tabular}{|ccccccc|}
\hline
model&${\Delta}z$&$\Omega_m$&$\Omega_\Lambda$&  $H_0$&$\epsilon$&$\epsilon/T_0$ \\
 SED   &      &        &           &        &   [Myr]  &   \\
\hline
GISSEL & 0.2  &  0.2   &  0.8      &   77.7 &   1695    &  0.12 \\
GISSEL & 0.3  &  0.2   &  0.8      &   71.5 &   1367    &  0.10 \\
PEGASE & 0.2  &  0.2   &  0.8      &   65.4 &   4101    &  0.30 \\
PEGASE & 0.3  &  0.2   &  0.8      &   53.0 &   2748    &  0.20 \\
\hline
\end{tabular}
\end{center}
\end{table*}
\begin{figure*}[!th]
\centerline{
\vbox{
\hbox{
\psfig{figure=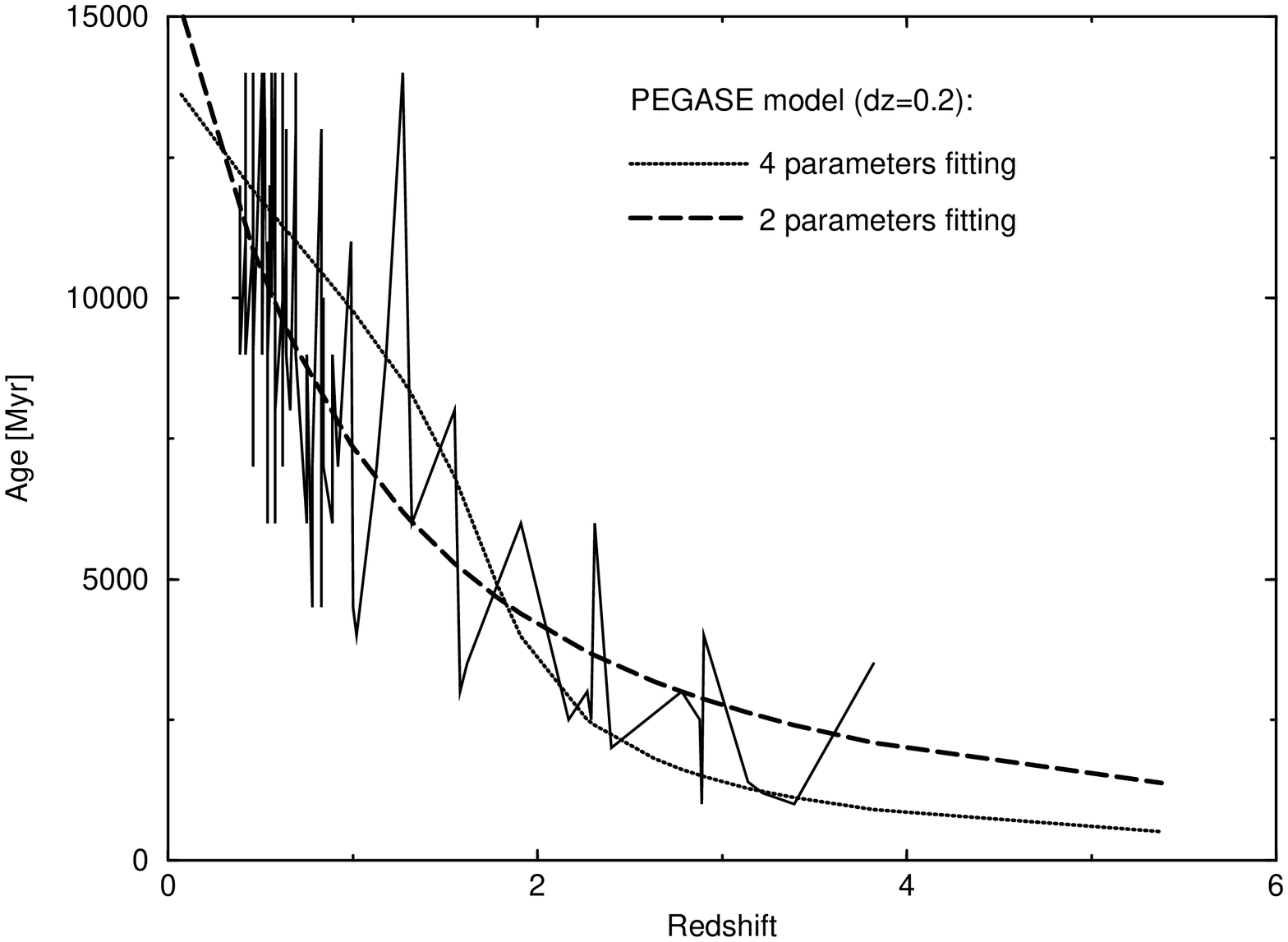,width=6cm,angle=-90}
\psfig{figure=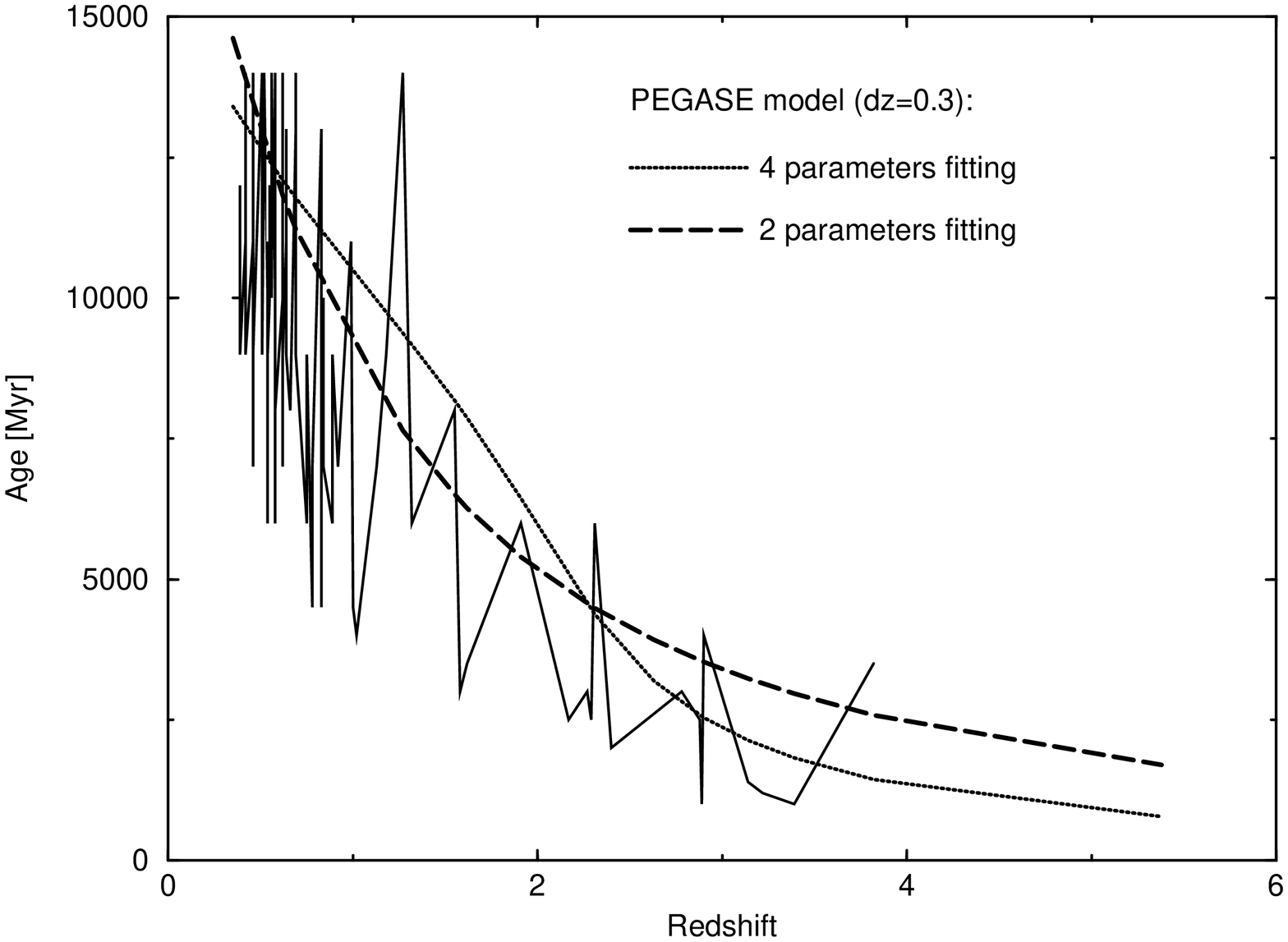,width=6cm,angle=-90}
}
\hbox{
\psfig{figure=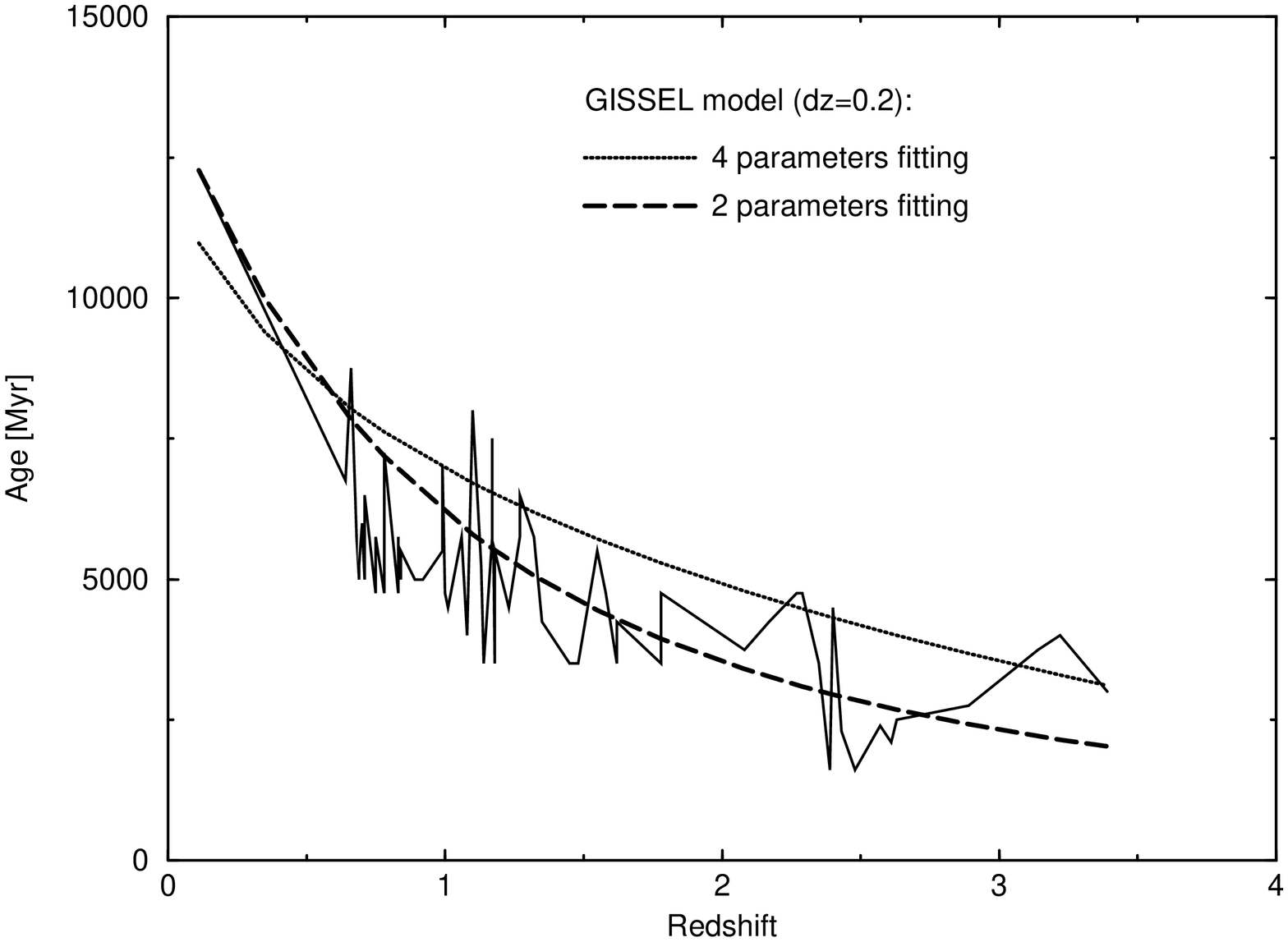,width=6cm,angle=-90}
\psfig{figure=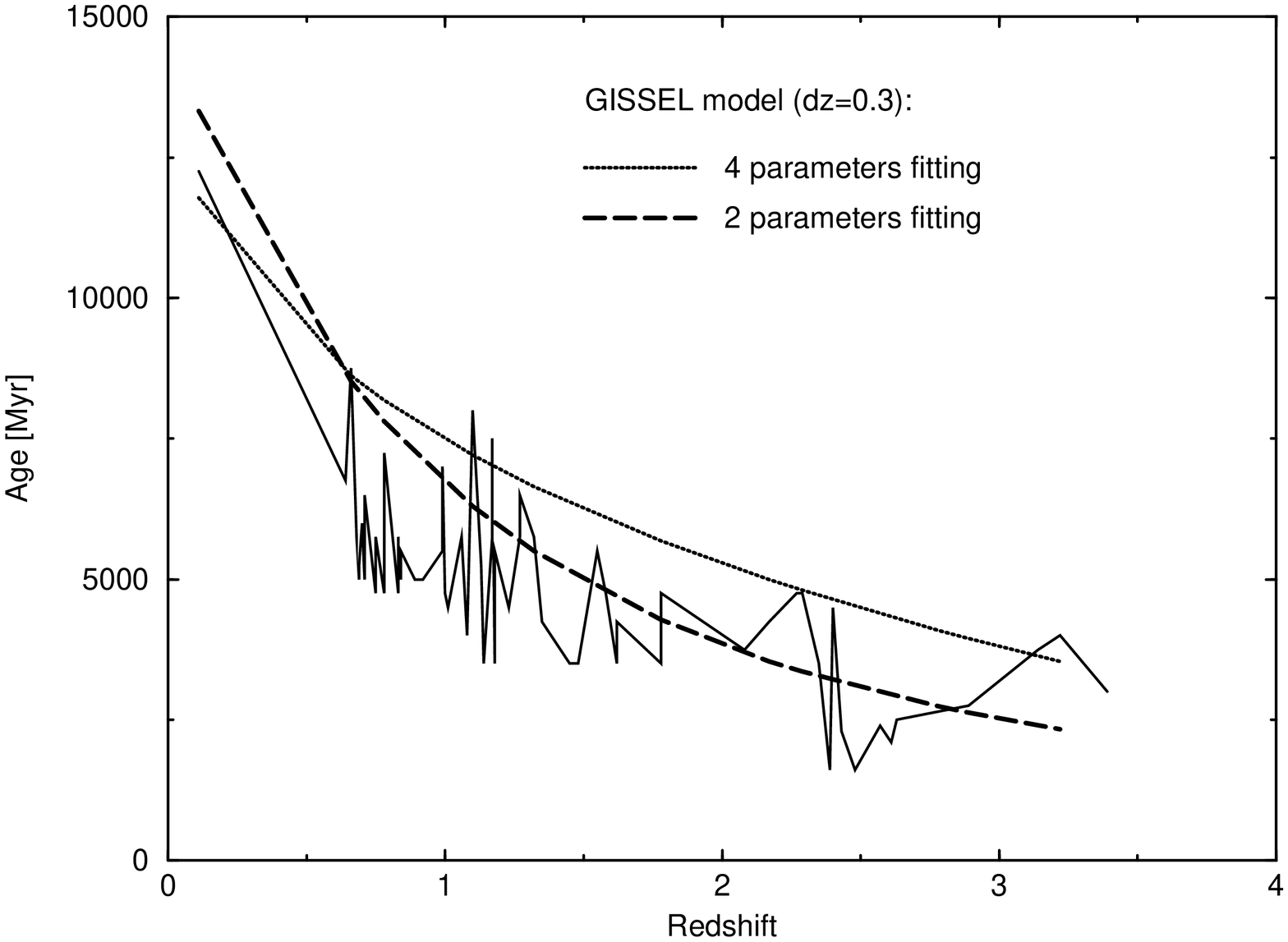,width=6cm,angle=-90}
}
}}
\caption{
The relationship $t(z)$ for models given in Table\,2.
Upper figures --- estimation for the model PEGASE, low figures ---
the models GISSEL.
Left figures --- discretization of $\Delta z$=0.2, right ones ---
discretization of $\Delta z$=0.3.
The curves are calculated for parameterization (5,6) using maxima of
galaxy ages in the given redshift intervals.
}
\end{figure*}

\subsection{Effects the errors have on the estimates of parameters}

This method of determination of $H_0$ and
$\Omega_\Lambda$ is stable enough concerning the input parameters
and systematic effect. As the modeling has shown the variation of the initial
metallicity caused a change in the age by 0.1\,Gyr (Jimenez \& Loeb 2002).
The change in the initial mass function does not affect the model SED either
(Bolzonella et al. 2000).

The error in the determination of the age, which may be connected with wrong
classification of the galaxy type and, therefore, with the choice of SED,
plays in this case in favor of the approach being described. The models SED
corresponding to elliptical galaxies give the oldest stellar population among
all the spectra. Thus, if in the galaxy are found star formation regions
giving a contribution to photometric data, then the choice of the optimum
model is displaced toward a younger stellar population, and the galaxy falls
out from our sample.

To check the influence of the sample completeness on the result, we used
the bootstrap method consisting in multiplying of
the initial sample and in constructing a new one by means of accidental
choice of objects. The coefficient of multiplying the sample was taken to be
equal to 100, and for each interval $\Delta z$ the number of objects equal to
the initial number was chosen in an accidental manner.
50 tests were thus applied,
and in each case the values of the parameters were estimated. As a result,
dispersion of the estimates $H_0=72\pm7$ for the model GISSEL and
$H_0=53\pm6$
for the model PEGASE for the interval $\Delta z=0.5$ were obtained.
We should note that accuracy of $\pm7$ for the Hubble parameter is the
inner accuracy of the method, and actually one elder galaxy than existing
in the sample can shift the estimate, but not far.

The values
of $\Omega_\Lambda$ remained unchanged, which is explained by the influence
of objects at moderate redshifts $z=0.3-1.0$ having a relatively
small dispersion of
ages. We disregarded the effect of the interval selfabsorption in the
galaxy (Sokolov et al. 2001) because of the ambiguity of solutions for a
small number of the used input parameters (number of filters) in our case
and the necessity for the determination of a large number of unknown
parameters.

The possible great contribution to the discrepancy in the approximation of
the relationships, which produces the effect of accidental errors, can be
considerably reduced by further statistical accumulation of data since the
number of known galaxies of type FR\,II will amount by different estimates to
a few thousand in the near future.

\section{Discussion of results}

The results of our paper confirm, firstly, that we live in an evolving
Universe. Secondly, that in contrast to the standard relativistic flat
model $\Lambda$CDM is situated within the errors of the method. To
estimate the quintessence $\omega_Q(z)$, the accuracy is not sufficient yet.
For the united data of different populations of elliptical galaxies,
for radio
galaxies too, an analysis was made of the upper limit of the age of formation
of stellar systems. From these data boundaries of determination of the
cosmological parameters $H_0$ and $\Lambda$--term were estimated:
$H_0=53\pm10$, and $\Omega_\Lambda=0.6\pm0.1$ in the model PEGASE. Note
that the models GISSEL have lower dispersion of ages for each interval, i.e.
they give a more stable result and, possibly, as a consequence, more reliable
cosmological parameters. The discrepancies of the relationship $t(z)$
decrease
and, therefore, the accuracy of estimates of parameters, when changing
from the interval $\Delta z=0.2$ to the interval $\Delta z=0.3$, improves,
which is explained by more reliable determination of the maximum age at a
larger interval.

As far as the procedure applied is concerned, it should be
noted that one of the main problems is the use of investigations
 of radio galaxies in which,
apart from standard evolution of stars, the photometric measurements may be
affected by other factors as well. Nevertheless, new models (for instance,
PEGASE2: Le Borgue \& Rocca-Volmerange 2002)
which allow these factors to be
taken into account, are already beginning to appear.

Note also that in principle

1) use of age characteristics of galaxies
for independent estimates of cosmological parameters is prospective;
example --- the estimate of the $\Lambda$--term, which can be improved
by extending the sample and employing more refined models;

2) the first examinations of the test ``Age of radio galaxies --- redshift''
yielded estimates close to the most accurate measurements of parameters from
the WMAP satellite data (Spergel et al., 2003).

3) preliminary selection of candidates must be performed by different ways
since none of the known ones is ideal. The relatively powerful radiation
(the ratio of radio luminosity to optical) points to the fact that we deal
with a giant galaxy with a supermassive black hole in the center, which
requires time to be formed. Unfortunately, there is no generally accepted
theory of their formation. The proposal to use objects in well understood
clusters also seems to be attractive (Kopylov 2001).
Because it is exactly from
clusters that data for $R(t)$ at small redshifts were obtained. The use of
the upper values of the age in the samples at different redshifts seems to be
justified. Even one the oldest object in the sample is decisive in the
determination of the lower limit of the Universe age at the given redshift,
similar to that as the only old star in the Galaxy (or the oldest globular
cluster) defines the minimum age of the Universe today. In our opinion the
most involved problem is the problem of the theory of evolution of the
distribution of energy in the spectrum of galaxies, and here we have
considerably  various readings.

\begin{acknowledgements}
OVV expresses his gratitude to RFBR for partial support of the work through
grant No 02--7--90038 and YuNP for support through grants of ``Integration'',
``Astronomy'' and RFBR. Special thanks are due to A.\,Kopylov for numerous
critical remarks and proposals to use new data on color characteristics of
elliptical galaxies and clusters. AAS was partially  supported through
grants of RFBR 02--02--16817 and 00--15--96699 and also by the program of RAS
``Astronomy''. The authors are grateful to N.F.\,Vojkhanskaya for valuable
comments made during reading the paper.
\end{acknowledgements}


\begin{thebibliography}{}
\bibitem{}
   Afanasiev V.L., Dodonov S.N., Moiseev A.V., Verkhodanov O.V.,
     Kopylov A.I., Parijskij Yu.N., Soboleva N.S., Temirova A.V.,
     Zhelenkova O.P., Goss W.M. 2002,
     Prepr. No 139 SPb, St.Petersburg branch SAO, 1
\bibitem{}
   Arimoto N., Yoshii Y. 1987. \aaa, {\bf 179}, 23
\bibitem{}
   Bruzual G., Charlot S. 1993. Astrophys. J., {\bf 405}, 538
\bibitem{}
   Bruzual G., Charlot S. 1996,
	ano\-ny\-mo\-us@ftp://gemini.tuc.noao.edu/pub/charlot/bc96
\bibitem{}
   Bolzonella M., Miralles J.-M., Pell\'o R. 2000,
	Astron. Astroph., {\bf 363}, 476 (astro-ph/0003380)
\bibitem{}
   Chambers K., Charlot S. 1990. Astrophys. J. Lett., 1990, {\bf 348}, L1
\bibitem{}
   De Breuck C., van Breugel W., Stanford S.A., R\"ottgering H.,
	Miley G., Stern D. 2002, \aj, {\bf 123}, 637
\bibitem{}
   Fioc M., Rocca-Volmerange B. 1997,  \aaa, {\bf 326}, 950
\bibitem{}
   Friaca A.C.S., Terlevich R.J., 1998, \mnras, {\bf 298}, 399
\bibitem{}
   Efstathiou G.,  Moody S., Peacock J.A.
     Percival W.J., Baugh C., Bland-Hawthorn J., Bridges T., Cannon R.,
     Cole S., Colless M., Collins C., Couch  W.,
     Dalton G., de Propris R., Driver S.P.,
     Ellis R.S., Frenk C.S., Glazebrook K.,
     Jackson C., Lahav O., Lewis I., Lumsden S.,
     Maddox S., Norberg P., Peterson B.A.,
     Sutherland W., Taylor K. 2002.
      \mnras, {\bf 330}, 29.
\bibitem{}
  Jarvis M.J., Rawlings S., Eales S., Blundell K.M., Bunker A.J.,
     Croft S.; McLure R.J., Willott C.J. 2001.
     \mnras, {\bf 326}, 1585
\bibitem{}
   Jimenez R., Loeb A., 2002, \apj, {\bf 573}, 37 (astro-ph/0106145)
\bibitem{}
   Kopylov A.I.,
   Goss W.M., Parijskij Yu.N., Soboleva N.S., Temirova A.V.,
   Zhelenkova O.P., Vitkovskij Val.V., Naugolnaya M.N., Verkhodanov O.V. 1995.
    \azh, {\bf 72}, 613
\bibitem{}
   Kopylov A.I. 2001. Private communication.
\bibitem{}
   Le Borgne D.,  Rocca-Volmerange B. 2002,
	  \aaa, {\bf 386}, 446
\bibitem{}
   Lilly S.,  1987, \mnras, 1987, {\bf 229}, 573
\bibitem{}
   Lilly S., 1990, ``Evolution of the Universe'' (ed. Kron R.G.),
	Astron. Soc. Pacific, 344
\bibitem{}
   Leibundgut B. 2001.
	Ann. Rev. Astron. Astrophys., {\bf 39}, 67
\bibitem{}
   Magorrian J., Tremaine S., Richstone D., Bender R., Bower G., Dressler A.,
      Faber S.M., Gebhardt K., Green R., Grillmair C., Kormendy J., Lauer T.,
      1998,
      \aj, {\bf 115}, 2285
\bibitem{}
   Miller G.E., Scalo J.M. 1979, \aaa, {\bf 41}, 513
\bibitem{}
   Moy E., Rocca-Volmerange B., 2002, \aaa, {\bf 383}, 46

\bibitem{}
   Parijskij Yu.N. 2001.
      ``Current Topics in Astrofundamental Physics: the Cosmic Microwave
      Background'', Proc. NATO Advanced Study Inst.,
      (ed. Norma G. Sanchez)
      Kluwer Acad. Publish, 219

\bibitem{}
   Parijskij Yu.N., Bursov N.N., Lipovka N.M.,
      Soboleva N.S., Temirova A.V. 1991. Astron. Astrophys. Suppl. Ser.,
      {\bf 87}, 1

\bibitem{}
   Parijskij Yu.N., Bursov N.N., Lipovka N.M.,
      Soboleva N.S., Temirova A.V., Chepurnov A.V. 1992.
      Astron. Astrophys. Suppl. Ser.,
      {\bf 96}, 583
\bibitem{}
   Parijskij Yu. N., Goss W.M., Kopylov A.I.,
      Soboleva N.S., Temirova A.V., Verkhodanov O.V., Zhelenkova O.P.,
      Naugolnaya M.N. 1996. \bsao, No {\bf 40}, 5
\bibitem{}
   Parijskij Yu.N., Goss W.M., Kopylov A.I.,
      Soboleva N.S.,
      Temirova A.V., Verkhodanov O.V., Zhelenkova O.P. 2000a,
      Astron. Astrophys. Trans., {\bf 19}, 297
\bibitem{}
   Parijskij Yu.N.,
    Soboleva N.S., Kopylov A.I., Verkhodanov O.V.,
    Temirova A.V., Zhelenkova O.P., Winn J., Fletcher A., Berke B. 2000b,
     \pazh, {\bf 26}, 493
\bibitem{}
   Pedani M. 2003.
     New Astronomy, {\bf 8}, 805
\bibitem{}
   Pipino A., Matteucci F. 2004.
     \mnras, {\bf 347}, 968
\bibitem{}
   Reuland M., van Breugel W., R\"ottgering H., de Vries W., Stanford S.A.,
     Dey A., Lacy M., Bland-Hawthorn J., Dopita M., Miley G. 2003,
     \apj, {\bf 592}, 755
\bibitem{}
   Rawlings S., Lacy M., Blundell K.M., Eales S.A.,
      Bunker A.J., Garrington S.T. 1996,
      Nature, {\bf 383}, p.502
\bibitem{}
   Rocca-Volmerange B., Guiderdoni B. 1988, \aas, {\bf 75}, 93
\bibitem{}
   Saini Tarun Deep,  Raychaudhury Somak,
       Sahni  Varun, Starobinsky  A.A., 2000,
		Phys. Rev. Lett., {\bf 85}, 1162
\bibitem{}
   Sahni Varun, Starobinsky A.A. 2000,
     Internat. Journ. Modern Phys. D, {\bf 9}, 373
\bibitem{}
   Soboleva N.S.,
    Goss W.M., Verkhodanov O.V., Zhelenkova O.P., Temirova A.V.,
    Kopylov A.I., Parijskij Yu.N. 2000.
     \pazh, {\bf 26}, 723
\bibitem{}
 Sokolov V.V., Fatkhullin T.A., Castro-Tirado A.J., Fruchter A.S.,
    Komarova V.N., Kasimova E.R., Dodonov S.N., Afanasiev V.L.,
    Moiseev A.V. 2001.
    \aaa, {\bf 372}, 438
\bibitem{}
  Spergel D.N., Verde L., Peiris H.V., Komatsu E.,
     Nolta M.R., Bennett C.L.,
    Halpern M., Hinshaw G., Jarosik N., Kogut A., Limon M., Meyer S.S.,
    Page L., Tucker G.S., Weiland J.L., Wollack E., Wright E.L.). 2003.
    \apj, {\bf 48}, 175
    (astro-ph/0302209)
\bibitem{}
   Stanford S.A., Eisenhardt Peter R., Dickinson Mark ,
      Holden B.P., Roberto De Propris, 2002,
      \apjs, {\bf 142}, 153
      (astro-ph/0203498)
\bibitem{}
  Starobinsky A.A., Parijskij Yu.N., Verkhodanov O.V. 2004.
     Proc. Sternberg Astron. Instit., {\bf LXXV},
     Book of Abstr. of All Russian Astron. Conf. VAC-2004
     ``Horizons of Universe'' (in Russian), MSU, ISSN 0371-6769, 198

\newpage
\bibitem{}
   van Breugel W.J.M., De Breuck C., Stanford S.A.,
      Stern D., R\"ottgering H., Miley G.K. 1999,
      \apj, 1999, {\bf 518}, 61
\bibitem{}
   Verkhodanov O.V., 1996, \bsao, No {\bf 41}, 149
\bibitem{}
   Verkhodanov O.V., Kopylov A.I., Parijskij Yu.N.,
	Soboleva N.S., Temirova A.V. 1998a,
	in ``Modern problems of extragalctic astronomy'',
	Puschino, May 25--29, Puschino Sci. Center, 24
\bibitem{}
   Verkhodanov O.V., Kopylov A.I., Parijskij Yu.N.,
	Soboleva N.S., Temirova A.V., Zhelenkova O.P., 1998b,
	in ``Prospects of Astronomy and Astrophysics For the New
	Millennium''.
	Joint European and National Astronomical Meeting,
	JENAM'98. 7th Europ. \& 65th Ann. Czech Astron. Conf.,
	Prague, 9-12 Sept., 1998b,  p.302.
\bibitem{}
     Verkhodanov O.V., Kopylov A.I., Parijskij Yu.N., Soboleva N.S.,
	Temirova A.V. 1999, \bsao, No {\bf 48}, 41
	(astro-ph/9910559)
\bibitem{}
     Verkhodanov O.V.,  Kopylov A.I., Zhelenkova O.P., Verkhodanova N.V.,
	Chernenkov V.N., Parijskij Yu.N., Soboleva N.S., Temirova A.V. 2000,
	Atsron. Astrophys. Trans., {\bf 19}, No 3-4, 662,
	(astro-ph/9912359, http://sed.sao.ru)
\bibitem{}
   Verkhodanov O.V., Parijskij Yu.N., Soboleva N.S.,
     Kopylov A.I., Temirova A.V., Zhelenkova O.P., W.M.Goss. 2002,
     \bsao, No {\bf 52}, 5 (astro-ph/0203522)
\bibitem{}
   Verkhodanov O.V., Parijskij Yu. N. 2003.
     \bsao, No {\bf 55}, 66

\end{thebibliography}
\end{document}